\documentclass[showpacs,twocolumn,prb,eqsecnum]{revtex4}
\usepackage{amsmath}
\usepackage{epsfig,color}
\usepackage{graphicx}
\usepackage{dcolumn}
\usepackage{bm}

\newcommand{\beq}{\begin{equation}}
\newcommand{\eeq}{\end{equation}}
\newcommand{\bea}{\begin{eqnarray}}
\newcommand{\eea}{\end{eqnarray}}
\newcommand{\bwt}{\begin{widetext}}
\newcommand{\ewt}{\end{widetext}}

\begin{document}

\title{Universal and non-universal renormalizations in Fermi liquids}
\author{Andrey V. Chubukov$^{1}$ and Dmitrii L. Maslov$^{2}$}
\date{\today}

\begin{abstract}
We discuss an interplay between  the Fermi-liquid (FL) theory and  diagrammatic
perturbative approach to interacting Fermi systems.
 In the FL theory for
Galilean-invariant systems,  mass renormalization $m^*/m$ comes exclusively
from fermions 
at the Fermi surface. We show that in a diagrammatic
perturbation theory the same result for $m^*/m$ comes from fermions both at and away from the Fermi surface.
The equivalence of the FL and pertubative approaches
 is based on a particular relation between self-energy contributions
from high- and low-energy fermions.
 We argue that  care has to
be exercised in the renormalization group approach to a FL in
order not to miss the high-energy contribution to $m^*/m$. 
 As particular examples, we discuss $m^*/m$ and the quasiparticle residue $Z$ for 2D and 3D systems with both $SU(2)$ and $SU(N)$ symmetries,
and  with a short-range 
interaction. We derive  
 an expression for the anisotropic part of the Fermi-liquid vertex
 in the large-$N$ limit of the $SU(N)$ case.
\end{abstract}

\affiliation{
$^{1}$Department of Physics, University of Wisconsin-Madison, 1150 University
Ave., Madison, WI 53706-1390\\
$^{2}$Department of Physics, University of Florida, P. O. Box 118440, Gainesville, FL
32611-8440
}
\maketitle
\section{Introduction}
Despite its apparent simplicity, the  Landau Fermi Liquid (FL) theory is one of
the most non-trivial theories of interacting fermions.~\cite
{agd,pines,anderson,baym} It states that the linewidth of a  state near the
Fermi surface (FS)  is smaller than its energy, so that  the quasiparticle
propagator $G(\omega, \mathbf{p})$  has a well-defined pole at $\omega = p_F
(|{\bf p}|-p_F)/m^* + {\mathcal O}(|\mathbf{p}|-p_F)^2$,  where $p_F$ is the Fermi momentum.
It also states that the quasiparticle residue $Z$ and effective mass $m^*$
are  expressed in terms of an interaction vertex $\Gamma^\omega_
{\alpha\beta,\gamma\delta} (p,q)$, where $p\equiv (\omega,\mathbf{p})$ and $%
q\equiv (\omega^{\prime},\mathbf{q})$, 
with  one of the two \lq\lq four-momenta\rq\rq\/
 on the FS, e.g.,
 $p \equiv(0,\mathbf{p}_F)$, where $\mathbf{p}_F\equiv (%
\mathbf{p}/|\mathbf{p}|)p_F$.  [The vertex $\Gamma^\omega_{\alpha\beta,%
\gamma\delta}$ is obtained from a fully renormalized, antisymmetrized vertex 
$\Gamma_{\alpha\beta,\gamma\delta}(p,q;p_1,q_1)$ in the limit of zero
momentum- and vanishing energy transfer, i.e., for $|\mathbf{p}_1|=|\mathbf{p%
}|$, $|\mathbf{q}_1|=|\mathbf{q}|$, $\omega_1\to \omega$, and $%
\omega_1^{\prime}\to \omega^{\prime}$. ]  Finally, the FL theory states
that,  for a Galilean invariant system ($\mathbf{p}^2/2m$ dispersion for a
free particle), 
which is the only case considered 
 in this paper , the 
effective mass $m^*$ is expressed via $\Gamma^\omega_
{\alpha\beta,\gamma\delta}  (p,q)$ with \textit{both} four-momenta on the
FS.  On the other hand, renormalization of the quasiparticle residue $Z$ 
 comes from fermions with $p = p_F$ but $q$,  in general, is 
 away from the FS.

Explicitly, for a Galilean-invariant system,~\cite{agd,pit} 
\begin{equation}
G_p = \frac{Z}{\omega - p_F (|\mathbf{p}|-p_F)/m^*},  \label{new_1}
\end{equation}
where 
\begin{subequations}
\begin{equation}
\frac{1}{Z} = 1 - \frac{i}{2} \sum_{\alpha\beta} \int
\Gamma^\omega_{\alpha\beta,\alpha\beta} (p_F,q) \left(G^2_q\right)_\omega 
\frac{d^{D+1} q}{(2\pi)^{D+1}},  \label{1_a}
\end{equation}
\begin{equation}
\frac{1}{m^*} = \frac{1}{m} - A_D \sum_{\alpha\beta} \int
\Gamma^\omega_{\alpha\beta,\alpha\beta} (p_F,q_F) \frac{\mathbf{p}_F\cdot 
\mathbf{q}_F}{p^2_F}~ d \Omega_q,  \label{1_b}
\end{equation}
$\Omega_q$ is the solid angle, $A_D = Z^2 k^{D-2}_F/2 (2\pi)^D$, $G_q$ is
the full fermionic propagator, and $\left(G^2_q\right)_\omega$ is the
product of the two Green's functions with the same momenta and
infinitesimally close frequencies. Note that integration in Eq.~(\ref{1_b})
is only over 
$d\Omega_q$, 
which implies that mass renormalization comes solely
from fermions  on the FS. In the field-theoretical language, mass renormalization
is then a low-energy, universal phenomenon, while a reduction of $Z$ from
its bare value of one is a high-energy,  non-universal phenomenon.

The effective mass $m^*$ and $Z$ factor can also be obtained 
by  expanding the self-energy $\Sigma (\omega, \epsilon_{\mathbf{p}})$ to
first order in $\omega$ and $\epsilon_{\mathbf{p}}$: 
\end{subequations}
\begin{equation}
\Sigma (\omega, \epsilon_{\mathbf{p}}) = \left (\omega - \epsilon_{\mathbf{p}%
}\right) \left(\frac{1}{Z}-1\right) - \epsilon_{\mathbf{p}} \left(\frac{m}{%
m^*}-1\right) + {\mathcal O}\left(\omega^2, \epsilon^2_p\right).  \label{a}
\end{equation}
[We define $\Sigma$ by $G_p^{-1} = \omega - \epsilon_{\mathbf{p}} + \Sigma
(\omega, \epsilon_{\mathbf{p}})$ with $\epsilon_{\mathbf{p}} = (\mathbf{p}%
^2-p^2_F)/2m$.] 
As, in practice, the self-energy is obtained via a diagrammatic perturbation
theory, we will refer to this approach as to \lq\lq perturbative\rq\rq\/. 
 In the earlier days of the FL theory, 
 perturbative calculations were used as a check of the general FL relations,
e.g., it has been verified~\cite{abrikosov58,galitskii} that 
 the \textit{values} of $m^*/m$ and $Z$ in Eq.~(\ref{a}) are the same as in
Eqs.~(\ref{1_a}) and (\ref{1_b}).
 However, whether mass renormalization in Eq.~(\ref{a}) comes from low
energies, as it does in Eq.~(\ref{1_b}), has not been verified.

In this paper we demonstrate that, in a diagrammatic calculation,  mass
renormalization \textit{is not}, in general,  a low-energy phenomenon. A
low-energy contribution to $m^*$ does, indeed, exists, but there is also
another, high-energy contribution.  Only the sum of the two contributions
reproduces the Landau formula for the effective mass, Eq.~(\ref{1_a}). 
 There are situations (see below) when the high-energy contribution is relatively small but, in general, it is of the same order as the low-energy one. 

The reason why low-energy mass renormalization is generally not the full
result in 
a diagrammatic calculation, can be traced back to 
 the fact that the building block  of diagrammatics is a
non-antisymmetrized interaction potential $U({\bf k})$ rather than the 
antisymmetrized vertex function $\Gamma^\omega$.  An expression for $%
\Gamma^\omega$ in terms of $U(|{\bf k}|)$ does contain  a high-energy
contribution and, when the self-energy is expressed  in terms of $%
U(|{\bf k}|)$ rather than in terms of $\Gamma^\omega$,  these high-energy terms  do contribute to the effective mass. When one re-expresses $\Sigma$ in terms
of $\Gamma^\omega$, the high-energy contributions to $m^*$ cancel out.

An issue where mass renormalization comes from is important for the
interpretation of a Fermi liquid as a fixed point of the 
 momentum-space renormalization group (RG) transformation.~\cite{shankar} 
In the RG approach, one progressively integrates out high-energy fermions ending up with a renormalized interaction among low-energy  ones. According to Eq. (%
\ref{1_b}), this interaction  is all one needs to evaluate the effective
mass.  Our finding that, in a diagrammatic calculation, $m^*/m$  may have
contributions from both low- and high-energy fermions implies that care has
to be exercised in applying the RG approach to a FL. 
Specifically,  
 we argue that to  recover the Landau formula for $m^*/m
$, one also  has to take into account  that, in the process of RG flow,  the
``bare'' mass for the low-energy theory changes from the free-fermion mass $m
$ to a different value ($m_B$). The difference $m_B/m -1$ comes from high
energies.  Only the sum of regular renormalization from $m$ to $m_B$ and 
low-energy renormalization of $m_B$ yields the agreement with the FL theory.

We further demonstrate that 
that there  exists a certain identity 
 [cf. Eq.~(\ref{10})], which 
relates the
high- and low-energy terms. 
 This
identity involves
combinations of  fermionic Green's functions 
in particle-hole and particle-particle channels, and is    
exact to first order in $\omega$ and $\epsilon_{\mathbf{p}}$. 
  Adding this identity to the diagrammatic 
self-energy $\Sigma (\omega, \epsilon_{\mathbf{p}})$
does not  change $%
{\mathcal O}(\omega)$ and ${\mathcal O}(\epsilon_{\mathbf{p}})$ terms in $\Sigma$,  i.e., it
does not
change $Z$ and $m^*/m$, but, at the same time, it 
  transforms  the high-energy contribution  into the low-energy one, and  makes
 the diagrammatic self-energy equivalent to the self-energy extracted from the 
FL theory. 
 
An interesting example of comparison between the FL and perturbative approaches
is the large-$N$ limit for an $SU(N)$-invariant 2D system with short-range interaction. 
The perturbative self-energy in this case is obtained simply by retaining the Random-Phase Approximation (RPA) diagrams with maximal
number of particle-hole bubbles at each order in the interaction. It is not enough, however, to retain only diagrams with a maximal
number of bubbles in order to construct $\Gamma^{\omega}$ because these diagrams contribute only to an isotropic part of $\Gamma^{\omega}$
and, therefore, do not lead to mass renormalization. We show that the perturbation theory for an anisotropic part of $\Gamma^{\omega}$ can be resummed to infinite order in $U$ even for subleading in $1/N$ terms, and the resulting expression for $m^*$ coincides with that obtained from the self-energy.

The structure of this paper is  as follows. In Sec.~\ref{flt}, we discuss
the FL theory and perturbation series for $\Gamma^\omega$. We briefly
discuss the 3D case, and  present the FL expressions for $m^*/m$ and $Z$ in
2D with a short-range interaction (to the best of our knowledge, the result for $Z$ has not been derived in the prior
literature.)  In Sec.~\ref{pert}, we obtain the self-energy in the
diagrammatic perturbation theory both in 3D and 2D  and identify the low-
and high-energy contributions to the effective mass.  We show that $m^*/m$
ad $Z$ are indeed the same as in the FL theory, but 
 at least part of 
 mass renormalization
comes from 
 high energies. Moreover, we show
that, in 2D, entire mass renormalization to second order in the interaction
comes from high energies, if the calculation is performed by 
combining internal fermions
into particle-hole pairs, 
while the high-energy part is twice larger and of opposite sign to the low-energy part,
 if internal fermions are combined into
particle-particle pairs. 
 In Sec.~\ref{recons}, we reconcile the two
approaches by proving  a particular relation between the convolutions of
Green's functions. 
 In Sec.~\ref{largeN} we discuss an extension of our results to the $SU(N)$
case 
 and consider  the large $N$ limit. 
 Finally,  
 in Sec.~\ref{conc} we present our conclusions.

\section{Fermi-liquid theory}

\label{flt}

\subsection{Pitaevskii-Landau relations}

We remind the reader that Eqs. (\ref{1_a}) and (\ref{1_b})  in the FL theory
are based on  the Pitaevskii-Landau relations -- the three identities for
the derivatives of the Green's function:\cite{agd,pit} 
\begin{widetext}
\bea
&&\frac{\partial G_p^{-1}}{\partial \omega} = \frac{1}{Z} = 1 - \frac{i}{2}  \sum_{\alpha\beta} 
\int \Gamma^\omega_{\alpha\beta,\alpha\beta}  (p_F,q) (G^2_q)_\omega 
\frac{d^{D+1} q}{(2\pi)^{D+1}}, \label{2_a} \\
&&{\bf p}_F\frac{\partial G^{-1}_p}{\partial {\bf p}} = - \frac{p^2_F}{m^*Z} 
 = -\frac{p^2_F}{m} + \frac{i}{2}  \sum_{\alpha\beta} 
\int \Gamma^k_{\alpha\beta,\alpha\beta}  (p_F,q) \frac{{\bf p}_F\cdot {\bf q}}{m} 
(G^2_q)_k \frac{d^{D+1} q}{(2\pi)^{D+1}},  \label{2_b} \\
&&\frac{1}{Z} = 1 - \frac{i}{2}  \sum_{\alpha\beta} 
\int \Gamma^\omega_{\alpha\beta,\alpha\beta}  (p_F,q) (G^2_q)_\omega \frac{ 
{\bf p}_F \cdot{\bf q}}{p^2_F}
\frac{d^{D+1} q}{(2\pi)^{D+1}}.
\label{2_c}
\eea
\end{widetext}
The first two relations originate from particle-number conservation, 
 while the third relation is a consequence  of Galilean invariance.  In Eq.~(%
\ref{2_b}), the object $\left(G^2_q\right)_k$ is  the product of two Green's
functions with the same frequencies and infinitesimally close momenta,  and $%
\Gamma^k_{\alpha\beta,
\gamma\delta
}$ is the vertex  in the limit of zero
frequency transfer and vanishing momentum transfer. The latter is related to 
$\Gamma^\omega_{\alpha\beta,
\gamma\delta
}$ by an integral equation 
\begin{eqnarray}
&&\Gamma^k_{\alpha \beta,\alpha\beta} (p,q) = \Gamma^\omega_{\alpha
\beta,\alpha\beta} (p,q)  \notag \\
&& -\frac{k^ {D-1}_F Z^2}{v_F (2\pi)^D}\sum_{\xi,\eta} \int
\Gamma^\omega_{\alpha \xi,\alpha\eta} (p,q^{\prime}) \Gamma^k_{\eta
\beta,\xi\beta} (q^{\prime},q) d \Omega_{q^{\prime}}.  \label{n_1}
\end{eqnarray}
In addition, $\left(G^2_q\right)_k$ is related to $\left(G^2_q\right)_\omega$
by 
\begin{equation}
\left(G^2_q\right)_k -\left(G^2_q\right)_\omega\equiv \delta G^2_q= - \frac{%
2\pi i Z^2 m^*}{p_F} \delta(\omega) \delta(|\mathbf{q}| -p_F).  \label{n_2}
\end{equation}
Note that Eqs.~(\ref{2_a}) and (\ref{2_b}) 
 contain the integrals over all intermediate states with momenta $q$.
However, using the additional property of Galilean invariance
 (\ref{2_c}), one can eliminate the high-energy contribution to $m^*$
(but not to $Z$). 
 Indeed, substituting Eqs.~(\ref{n_1}) and (\ref{n_2}) into Eq.~(\ref{2_b}%
) and using Eq.~(\ref{2_c}), one reduces Eq.~(\ref{2_b}) to 
\begin{eqnarray}  \label{2_bb}
&&\mathbf{p}_F\frac{\partial G^{-1}_p}{\partial \mathbf{p}} = - \frac{p^2_F}{%
m^*Z} = -\frac{p^2_F}{mZ} + \frac{i}{2} \frac{p^2_F}{m^*Z}  \notag \\
&& \times\sum_{\alpha\beta} \int \Gamma^\omega_{\alpha\beta,\alpha\beta}
(p_F,q) \delta G^2_q \frac{\mathbf{p}_F \cdot\mathbf{q}}{p^2_F} \frac{%
d^{D+1} q}{(2\pi)^{D+1}}  \notag \\
&&= -\frac{p^2_F}{mZ} + \frac{p^2_F}{Z} A_D \sum_{\alpha\beta} \int
\Gamma^\omega_{\alpha\beta,\alpha\beta} (p_F,q_F) \frac{\mathbf{p}_F\cdot 
\mathbf{q}_F}{p^2_F}~ d \Omega_q,  \notag \\
\end{eqnarray}
which is equivalent to Eq. (\ref{1_b}) for mass renormalization. We
emphasize again that Eq.~(\ref{1_b}), which involves only low-energy
fermions, is based not only on 
 particle-number conservation [Eqs. (\ref{n_1}) and (\ref{n_2})], but  also on Eq.~(\ref{2_c}), specific
only  for Gallilean-invariant systems.

Combining Eqs. (\ref{2_a}), (\ref{2_bb}), and (\ref{2_c}),  one  
 can construct the
self-energy to first order in $\omega$ and $\epsilon_{\mathbf{p}}$ as 
\begin{widetext}
\beq
\Sigma_{\mathrm{FL}} (\omega, \epsilon_{\mathbf{p}}) = \left(\omega -\epsilon_{\mathbf{p}}\right) 
\left[-\frac{i}{2}  \sum_{\alpha\beta} 
\int \Gamma^\omega_{\alpha\beta,\alpha\beta}  (p_F,q) G^2_q 
\frac{d^{D+1} q}{(2\pi)^{D+1}}\right] +  \epsilon_{\mathbf{p}} \left[\frac{i}{2Z}  \sum_{\alpha\beta} 
\int \Gamma^\omega_{\alpha\beta,\alpha\beta}  (p_F,q) \frac{{\bf p}_F \cdot{\bf q}}{p^2_F} \delta G^2_q  
\frac{d^{D+1} q}{(2\pi)^{D+1}}\right].
\label{15}
\eeq
\end{widetext} 

\subsection{ Perturbation theory for $\Gamma^\protect\omega$}

\begin{figure}[tbp]
\centering
\includegraphics[width=1.0\columnwidth]{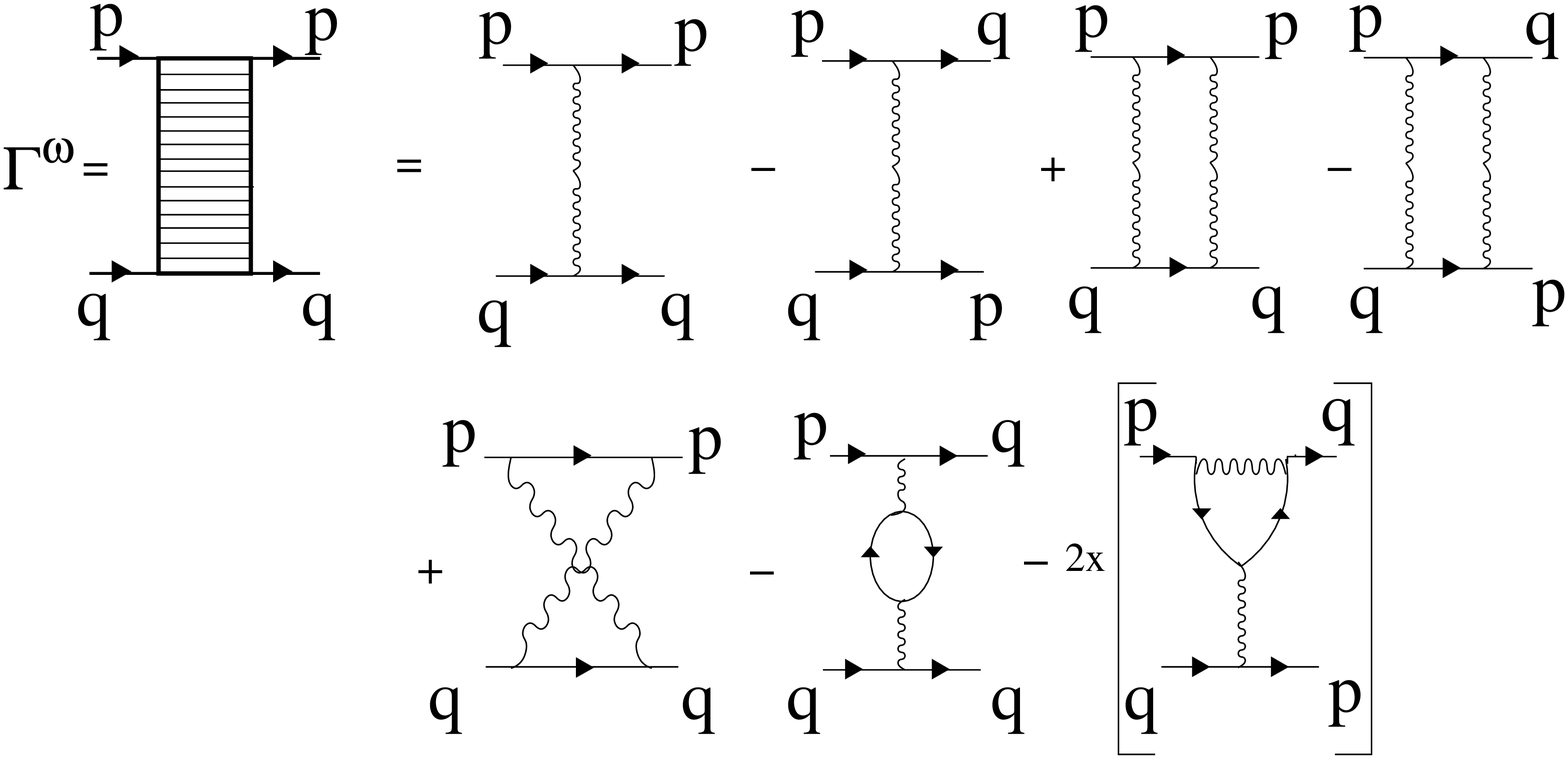}
\caption{ First and second order diagrams for the Fermi-liquid vertex $%
\Gamma^{\protect\omega}_{\protect\alpha\protect\beta,\protect\gamma\protect%
\delta}(p,q)$. The initial four-momenta $p$ and $q$ are associated with spin
projections $\protect\alpha$ and $\protect\beta$, respectively. The final
four-momenta $p$ and $q$ are associated with spin projections $\protect\gamma
$ and $\protect\delta$, respectively.}
\label{fig:fig1}
\end{figure}

The vertex $\Gamma^\omega$ can be obtained  via a perturbative expansion in $%
U(|{\bf k}|)$. Diagrams for $\Gamma^\omega$ to second order in $U(|{\bf k}|)$ are presented
in Fig.~\ref{fig:fig1}. Assume first that $U(|{\bf k}|) = \mathrm{const}\equiv U$
(contact interaction).  In this case, 
\begin{widetext}
\beq
\Gamma^\omega_{\alpha\beta,
\gamma\delta}  (p_F,q) = 
\delta_{\alpha\gamma}\delta_{\beta\delta} 
\left[U + i U^2 \int \left(G_l G_{q-p_F+l} + G_l G_{q+p_F-l}\right)
  \frac{d^{D+1} l}{(2\pi)^{D+1}}\right] - 
 \delta_
 {\alpha\delta}\delta_{\beta\gamma} 
 \left[U + i U^2  \int  G_l G_{q+p_F-l}
  \frac{d^{D+1} l}{(2\pi)^{D+1}}\right].
\label{3}
\eeq
\end{widetext}
The first term  in Eq.~(\ref{3}) is the renormalized interaction with zero
momentum transfer, the second term is obtained by antisymmetrization.  We
see that the first (``direct'') term  contains contributions from both the
particle-hole and particle-particle channels, while the second
(``exchange'') term  contains only a contribution from the particle-particle
channel.

In 3D, explicit expression for $\Gamma^\omega$ when both particles are on
the FS (i.e., $q = q_F$) were obtained 
 long time ago (see Refs.~\onlinecite{abrikosov58,agd}). 
 A similar
calculation, performed in Ref.~\onlinecite{randeria} for the 2D case, yields 
\begin{eqnarray}
&&\Gamma^\omega_{\alpha\beta,\gamma\delta} (p_F,q_F)
=\Gamma^\omega_{\alpha\beta,\gamma\delta} (\theta)  \notag \\
&&=\frac{1}{2} \delta_{\alpha\gamma}\delta_{\beta\delta} \left[U + \frac{U^2
m}{\pi} +\frac{U^2 m}{2\pi} \ln\left(\cos\frac{\theta}{2}\right)\right] 
\notag \\
&& -\frac{1}{2} {\boldsymbol\sigma}_{\alpha\gamma} \cdot{\boldsymbol\sigma}%
_{\beta\delta} \left[U +\frac{U^2 m}{2\pi} \ln\left(\cos\frac{\theta}{2}%
\right) \right], \label{n_3}
\end{eqnarray}
where $\theta$ is the angle between $\mathbf{p}_F$ and $\mathbf{q}_F$. In
deriving Eq.~(\ref{n_3}), 
we used a result for the static particle-particle bubble in 2D 
\begin{eqnarray}
\Pi_{\mathrm{pp}}(\omega=0,|\mathbf{k}|\leq 2p_F)&=& i\int \frac{d^3 l}{%
(2\pi)^3} G_lG_{k-l}  \notag \\
&=& \frac{m}{2\pi}\ln\frac{2p_F}{|\mathbf{k}|}  \label{pic}
\end{eqnarray}
(up to an irrelevant constant). 
In 2D, the angular dependence of $%
\Gamma^{\omega} (\theta)$,  which is responsible for mass renormalization,
comes entirely from  the interaction in the particle-particle channel. Since
the 2D particle-hole bubble 
$\Pi_{\mathrm{ph}}(k)=i \int d^3lG_lG_{l+k}/(2\pi)^3$ is independent of $|\mathbf{k}|$ for $|\mathbf{k}|\leq 2p_F$, renormalization of the interaction in the particle-hole channel
only adds a constant to $U$ and is, therefore, irrelevant for $m^*$. The
formula for $\Gamma^\omega_{\alpha\beta,\gamma\delta} (p_F,q)$ for $q$  is
away from the FS is rather complex, and we refrain from presenting it.

\subsection{Effective mass and quasiparticle residue}

Substituting $\Gamma^\omega$ from Eq.~(\ref{3}) into Eqs.~(\ref{1_a}) and (%
\ref{1_b}) and evaluating the integrals in 3D, we reproduce the known
results for $m^*/m$ (Refs.~\onlinecite{abrikosov58,galitskii}) 
 \begin{equation}
\frac{m^*}{m} = 1 + \left(\frac{8}{15}\right) \left(7 \ln 2 -1 \right) \left(%
\frac{mUp_F}{4\pi^2}\right)^2  \label{4_a}
\end{equation}
and $Z$ (Ref.~\onlinecite{galitskii}) 
\begin{equation}
Z = 1 - 8 \ln 2 \left(\frac{mU p_F}{4\pi^2}\right)^2.  \label{4_aa}
\end{equation}
 In 2D, Eq.~(\ref{n_3}) immediately gives~\cite{randeria} 
\begin{equation}
\frac{m^*}{m} = 1 + \frac{1}{2} \left(\frac{mU}{2\pi}\right)^2,  \label{4-1}
\end{equation}
while for $Z$ we obtain
  after numerical integration of $\Gamma^\omega_{\alpha\beta,
\gamma\delta}(p_F,q)$ instead of (\ref{n_3}) 
\begin{equation}
Z \approx 1 - C\left(\frac{mU}{2\pi}\right)^2,  \label{4}
\end{equation}
where $C=0.6931...$. To 
 high numerical accuracy, 
 $C$ is equal to $\ln 2$, but we did not attempt to prove this analytically.
We remind that, in the FL formulation, $m^*/m$ comes exclusively from the
interaction between particles on the FS.

\subsubsection{Momentum-dependent interaction}

For a momentum-dependent interaction $U(|\mathbf{k}|)$,  expressions for $%
m^*/m$ and $Z$ are generally more complex. Mass renormalization now occurs
already at the first order in $U (|q|)$. In 3D, 
\begin{equation}
\frac{m^*}{m} = 1 - \frac{m p_F}{16 \pi^2}   \notag
\end{equation}
\begin{equation}
\times\int_0^2 d z z \left[U\left(p_F(2-z)^{1/2}\right) - U\left( p_F
(2+z)^{1/2}\right)\right] + O(U^2).  \label{n_4}
\end{equation}
Renormalization of $Z$ still occurs beginning from the second order in $U(|%
\mathbf{k}|)$.

\section{ Perturbation theory for the self-energy}

\label{pert}

\begin{figure}[tbp]
\centering
\includegraphics[width=1.0\columnwidth]{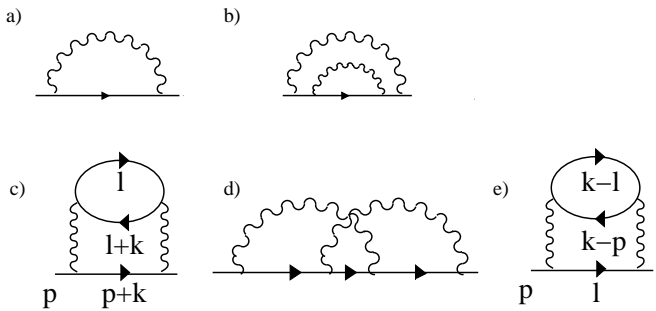}
\caption{ First and second order diagrams for the fermionic self-energy $%
\Sigma_{\mathrm{pert}} (\protect\omega, \protect\epsilon_{\mathbf{p}})$. For
a momentum-indepedent interaction $U(|{\bf q}|) =U$, only second-
 and higher-order diagrams renormalize the
mass and $Z$. For a momentum-dependent interaction, mass renormalization starts already at the first-order. Diagram {\it e} is the same as  {\it c}, except for internal fermions are combined into a particle-particle rather than a particle-hole pair.}
\label{fig:fig2}
\end{figure}

We now discuss how $m^*/m$ and $Z$ occur
 in the diagrammatic  perturbation
theory. Again, we consider first  the case of a constant interaction $U$;  a
momentum-dependent interaction will be discussed later.  The first-order
term in the self-energy (Fig.~\ref{fig:fig2}, {\it a})  is irrelevant in this case, and we focus on 
 the 
second-order diagrams in Fig.~\ref{fig:fig2}.  Diagram \textit{b}
 just shifts the chemical potential and is also irrelevant, so we need to consider
only diagrams \textit{c} and \textit{d}. Relabeling the fermionic momenta,
it is easy to see that,  for a constant $U$, 
 diagram \textit{d} is equal to $-1/2$ of
 diagram \textit{c}, so there is essentially one second-order diagram  to be considered, e.g.,
diagram \textit{c}. 
 This diagram contains three Green's functions, two of which share a common
internal momentum.  Labeling the momenta as shown in diagram \emph{c} and
integrating over the internal four-momentum $l$, we end up with a particle-hole
bubble.  Alternatively, labeling the momenta as shown in diagram \emph{e}
and  integrating over $l$, we end up with a particle-particle  bubble. 

We start with combining two internal fermions into a particle-hole bubble; 
the particle-particle combination is discussed in Sec. ~\ref{sec:pp}. 
Subtracting  from the particle-hole form of $\Sigma (\omega,\epsilon_{%
\mathbf{p}})$  its value at $\omega =0, \epsilon_{\mathbf{p}} =0$, we find 
\begin{equation}
\Sigma_{\mathrm{pert}} (\omega, \epsilon_{\mathbf{p}}) - \Sigma (0,0) = -U^2
\int G_l G_{k-p_F+l}~ \left(G_{k+\epsilon}-G_k\right) d_{lk},  \label{n_5}
\end{equation}
where $d_{lk} \equiv d^{D+1} l d^{D+1} k/(2\pi)^{2(D+1)}$ and 
\begin{equation}
\epsilon= \left(\omega, \frac{\epsilon_{\mathbf{p}}}{v_Fp_F}\mathbf{p}%
_F\right)  \label{eps}
\end{equation}
is the (small) external four-momentum. The self-energy can be further split
into two parts as 
\begin{equation}
\Sigma_{\mathrm{pert}} (\omega, \epsilon_{\mathbf{p}}) - \Sigma (0,0) =
\delta\Sigma_1 (\omega,\epsilon_{\mathbf{p}}) + \delta \Sigma_2
(\omega,\epsilon_{\mathbf{p}}),  \label{5}
\end{equation}
where 
\begin{subequations}
\begin{eqnarray}  
\delta \Sigma_1 (\omega,\epsilon_{\mathbf{p}}) &=& U^2 \int G_l G_{k-p_F+l}
G^2_q \left(\omega -\epsilon_{\mathbf{p}} \frac{\mathbf{p}_F \cdot\mathbf{k}%
}{p^2_F}\right) d_{lk}\notag\\
\label{6_a}\\
\delta\Sigma_2 (\omega,\epsilon_{\mathbf{p}}) &=& - U^2 \int^{\prime} G_l
G_{k-p_F+l}~ \left(G_{k-\epsilon}-G_k\right) d_{lk}.  \notag\\
\label{6_b}
\end{eqnarray}
\end{subequations}
The difference between the two parts is as follows. In the first part,  the
integrand was expanded to first order in $\omega$ and $\epsilon_{\mathbf{p}}$%
. This is justified if typical internal energies remain finite when $\omega,
\epsilon_{\mathbf{p}} \to 0$. This is a  regular, high-energy contribution
to the self-energy coming from fermions not confined to the FS.
  The second term
is an anomalous  contribution from internal  energies of order $\omega$ and $%
\epsilon_{\mathbf{p}}$,  which cannot be obtained by an expansion of $%
\Sigma_{\mathrm{pert}}$ in the external energies. 
 This second term is 
 a low-energy contribution (to  emphasize this, we put a prime on the
integral for this part).

\subsection{3D case}

Evaluating 
the
integrals in Eqs.~(\ref{6_a}) and (\ref{6_b}) for the 3D case, we
find that both contributions are finite; namely 
\begin{eqnarray}
\delta\Sigma_1 (\omega,\epsilon_{\mathbf{p}}) &=& 8 \ln 2 \left(\frac{mUp_F}{%
4 \pi^2}\right)^2 (\omega - \epsilon_{\mathbf{p}})  \notag \\
&&+\frac{4}{3} \epsilon_{\mathbf{p}} \left(4 \ln 2 -1\right) \left(\frac{%
mUp_F}{4 \pi^2}\right)^2,  \notag \\
\delta\Sigma_2 (\omega,\epsilon_{\mathbf{p}}) &=& -\frac{4}{5} \epsilon_{%
\mathbf{p}} \left(2 \ln 2 -1\right) \left(\frac{mUp_F}{4 \pi^2}\right)^2.
\label{7}
\end{eqnarray}
Adding up two parts and casting the result into the form of Eq.~(\ref{a}), 
we recover the Galitskii's result (Ref. \onlinecite{galitskii}): 
\bea
\Sigma_{\mathrm{pert}} (\omega, \epsilon_{\mathbf{p}}) &-& \Sigma (0,0) 
 = \left(\frac{mUp_F}{4 \pi^2}\right)^2  \label{new_a}\\
&&\times \left[8\ln 2~
(\omega - \epsilon_{\mathbf{p}}) + \frac{8}{15} \left(7 \ln 2-1\right)
 \epsilon_{\bf p}
\right] \notag
\eea
This self-energy indeed produces the same $m^*/m$ and $Z$ as in the FL
theory, Eqs.~(\ref{4_aa}) and (\ref{4_a}),  We see, however, that mass
renormalization--determined by a stand-alone $\epsilon_{\mathbf{p}}$ term in
the self-energy --comes from both  the high- and low-energy parts of $\Sigma_{%
\mathrm{pert}}$.  Only the sum of the two contributions recovers the  FL
formula for $m^*/m$. On the other hand,  renormalization of $Z$ comes only
from $\delta \Sigma_1$, i.e.,  from high energies.

\subsection{2D case}

The difference between the  FL and diagrammatic approaches becomes even more obvious in 2D. Since, to the best of our knowledge, the
 $Z$ factor for a 2D Fermi liquid with a short-range interaction has not been calculated before,  we consider 
  the 2D case in more detail. We also use 
  the
 2D case as an example to show that  an
  interplay between high-energy and low-energy contributions to the effective mass (but not the full result) 
 depends on whether the self-energy is calculated via particle-hole or particle-particle bubbles.

\subsubsection{perturbative self-energy  via particle-hole bubble}

The calculation of the perturbative self-energy via  a particle-hole bubble is based on Eqs.~(\ref{6_a}) and (\ref{6_b}). 
 It 2D, the particle-hole bubble can be found analytically for any $\omega$ and $|{\bf k}|$:
\begin{widetext}
\beq
\Pi_{\mathrm{ph}}(\omega ,{\bf k})=-\frac{m}{2\pi }\left[ 1+i\frac{\sqrt{2}{\tilde \omega} }{\sqrt{{\tilde k}^{2}-{\tilde k}^{4}-{\tilde \omega} ^{2}+\sqrt{\left({\tilde k}^{2}-{\tilde k}^{4}-{\tilde\omega} ^{2}\right) ^{2}-4{\tilde\omega} ^{2}{\tilde k}^{4}}}}\right],
\label{piph}
\eeq
\end{widetext}
where ${\tilde\omega}=2\omega m/p_F^2$ and ${\tilde k}=|{\bf k}|/2p_F$. 
To calculate the regular part of the self-energy, one needs to know the entire bubble, while the anomalous part is determined only by the static bubble $\Pi_{\mathrm{ph}}(0,|{\bf k}|)$. Performing the angular integral in $\delta\Sigma_1$ analytically and remaining integrals numerically, and all integrals in $\delta\Sigma_2$ analytically, we obtain
 \begin{subequations}
\begin{eqnarray}
\delta\Sigma_1 (\omega,\epsilon_{\mathbf{p}}) = && C \left(\frac{mU}{2\pi}%
\right)^2\!\!\! (\omega - \epsilon_{\mathbf{p}}) + \frac{\epsilon_{\mathbf{p}%
}}{2} \left(\frac{mU}{2 \pi}\right)^2  \label{8-1} \\
\delta\Sigma_2 (\omega,\epsilon_{\mathbf{p}}) = &&0,  \label{8}
\end{eqnarray}
with 
\end{subequations}
\begin{eqnarray}
C=\frac{1}{2\pi}\int_{0}^{\infty }\!\!dx\int_{0}^{\infty }\!\!dy \frac{%
\partial }{\partial y }\frac{\sqrt{A+\sqrt{A^{2}+x^2y^2}}}{\sqrt{A^{2}+x^2y^2%
}}F\left( x ,y\right),  \notag \\
\end{eqnarray}
$A =\left(x-x^2+y ^{2}\right)/2$, and 
\begin{eqnarray}
F(x,y)= \frac{y }{\sqrt{A+\sqrt{A^{2}+x^2y^2}}}-\theta \left( x-1\right) 
\sqrt{\frac{x-1}{x}}.
\end{eqnarray}
Numerical integration yields, 
 to high accuracy,  
 $C = 0.6931\dots = \ln{2}$; same as for the prefactor in Eq.~(\ref{4}).

 The reason why $%
\delta\Sigma_2=0$ in 2D is very simple. 
This contribution is expressed via a static particle-hole bubble as 
\begin{eqnarray}
\delta\Sigma_2 (\omega,\epsilon_{\mathbf{p}}) &=& i\frac{\epsilon_{\mathbf{p}} U^2}{2\pi^2 v_F}%
\int^{2p_F}_0d|\mathbf{k}| \Pi_{\mathrm{ph}}(\omega=0,|\mathbf{k}|)  \notag
\\
&&\times\frac{1-|\mathbf{k}|^2/2p_F^2}{\sqrt{1-\left(|\mathbf{k}%
|/2p_F\right)^2}}.  \label{s2}
\end{eqnarray}
As $\Pi_{\mathrm{ph}}(\omega=0,|\mathbf{k}|)$ is independent of $|\mathbf{k}|
$ for $\mathbf{k}|\leq 2p_F$, the integral over $|\mathbf{k}|$ vanishes. For
the same reason, mass renormalization in the FL theory comes only from the
particle-particle part of the vertex in Eq.~(\ref{n_3}).

Casting the result into the form of Eq.~(\ref{a}), we again reproduce the FL
results for $m^*/m$ and $Z$, Eqs.~(\ref{4-1}) and (\ref{4}). However, we see
that now $m^*/m$ comes solely from the  high-energy part of the self-energy,
in an (apparent) contradiction to the FL theory, where it comes  from low
energies.

\subsubsection{perturbative self-energy via particle-particle bubble}
\label{sec:pp} 

We now show that an
interplay between high- and low-energy contributions to the perturbative self-energy depends on the way how we obtain it. To demonstrate this, 
 we 
 obtain
 the 
same
 perturbative self-energy $\Sigma_{\mathrm {pert}} (\omega, \epsilon_p)$
 as in Eq.~(\ref{8-1})
  by combining internal fermions into particle-particle rather than particle-hole pairs. We will see
 that in this situation mass renormalization comes 
  from both high- and low energies.
 
For simplicity, we consider a momentum-independent interaction in 2D and
restrict attention to the second-order in $U$.
Labeling the momenta as shown in diagram \emph{e} of Fig.~\ref{fig:fig2}, we
obtain for the second-order self-energy 
\begin{equation}
{\tilde \Sigma}_{\mathrm{pert}} (\omega, \epsilon_{\mathbf{p}}) - {\tilde
\Sigma} (0,0) = -U^2 \int G_l G_{k+p_F-l}~ \left(G_{k-\epsilon}-G_k\right)
d_{lk}.  \label{n_55}
\end{equation}
We denote the self-energy obtained in this way as ${\tilde \Sigma}_{\mathrm{%
pert}}$ to distinguish it from the self-energy in the particle-hole form. 
 As before, we split the difference 
${\tilde \Sigma}_{\mathrm{pert}} (\omega,\epsilon_{\mathbf{p}}) - {\tilde \Sigma} (0,0)$ into a sum of regular and anomalous contributions as 
\begin{subequations}
\begin{eqnarray}  \label{66_b}
\delta {\tilde \Sigma}_1 (\omega,\epsilon_{\mathbf{p}}) &=& U^2 \int G_l
G_{k+p_F-l} G^2_k \left(\omega -\epsilon_{\mathbf{p}} \frac{\mathbf{p}_F
\cdot\mathbf{k}}{p^2_F}\right) d_{lk}  \notag \\
\\
\delta{\tilde \Sigma}_2 (\omega,\epsilon_{\mathbf{p}}) &=& -U^2
\int^{\prime} G_l G_{k+p_F-l}~ \left(G_{k-\epsilon}-G_k\right) d_{lk}. 
\notag \\
\end{eqnarray}
\end{subequations}
 The anomalous part $\delta{\tilde \Sigma}_2$ is now expressed via a static 
\emph{particle-particle} bubble as
\begin{eqnarray}
\delta{\tilde\Sigma}_2(\omega,\epsilon_{\mathbf{p}})&=&-\frac{\epsilon_{%
\mathbf{p}} U^2}{2 \pi^2 v_F}\int^{2p_F}_0d|\mathbf{k}| \Pi_{\mathrm{pp}%
}(\omega=0,|\mathbf{k}|)  \notag \\
&&\times\frac{1-|\mathbf{k}|^2/2p_F^2}{\sqrt{1-\left(|\mathbf{k}%
|/2p_F\right)^2}},  \label{s2pp}
\end{eqnarray}
where $\Pi_{\mathrm{pp}}(\omega=0,|\mathbf{k}|)$ is given in Eq.~(\ref{pic}%
). In contrast to Eq.~(\ref{s2}), the integral over $|\mathbf{k}|$ now does
not vanish,  and we obtain 
\begin{equation}
\delta{\tilde\Sigma}_2(\omega,\epsilon_{\mathbf{p}})= -\frac{\epsilon_{%
\mathbf{p}}}{2}\left(\frac{mU}{2\pi}\right)^2.  \label{new_n2}
\end{equation}
Therefore, in contrast to the particle-hole case, the low-energy
contribution to mass renormalization in the particle-particle case is finite
but opposite in sign to mass renormalization in the FL theory, Eq.~(\ref{4-1}%
). 

The regular part of self-energy, $\delta{\tilde\Sigma}_1$, renormalizes
both the $Z$ factor and effective mass. To evaluate this 
 part of the
 self-energy, a
static approximation for $\Pi_{pp}$  is not sufficient, and we need 
 a dynamic form of the particle-particle propagator. Re-expressing 
  $\delta{\tilde\Sigma}_1$ in (\ref{66_b}) via  
    the  particle-particle polarization 
   bubble
  and shifting 
  the
  momentum as $k-p_F \to k$, we obtain
\begin{eqnarray}
&&\delta {\tilde \Sigma}_1 (\omega,\epsilon_{\mathbf{p}}) = U^2
 \int \frac{\Pi_{pp} (\omega_k, {\bf k})}{(\omega_k - \epsilon_{k-p_F})^2}
\nonumber \\
&& \times
 \left[(\omega -\epsilon_{\mathbf{p}}) + \epsilon_{\mathbf{p}} \left(2 -
 \frac{\mathbf{p}_F \cdot\mathbf{k}}{p^2_F}\right)\right] d_{k}.
\label{l_1}
\end{eqnarray}
We remind that $d_k =  k dk d \theta_k d\omega_k/(2\pi)^3$.  If we approximated $\Pi_{pp} (\omega_k, {\bf k})$ by its satic form,  $\delta{\tilde\Sigma}_1$ would vanish 
 after integration over $\omega_k$ because of the double pole. The 
 dynamic $\Pi_{pp} (\omega_k, {\bf k})$,
 however, has branch cuts in both upper and lower half-planes of  $\omega_k$,  which ensures that the frequency integral is non-zero.

The dynamic polarization bubble is obtained by standard means and, for $|{\bf k}| < 2p_F$, is given by
\begin{equation}
\Pi_{{\mathrm pp}} (\omega_k, {\bf k}) = \frac{m}{2\pi^2} \int_0^{\pi/2} d \phi 
\ln{S_\phi (\omega_k, {\bf k})}
\label{l_2}
\end{equation}
where
\begin{eqnarray}
&&S_\phi (\omega_k, {\bf k}) = \frac{m \omega_k - p^2_F +{\bf k}^2/4}{m \omega_k
 -|{\bf k}| \cos{\phi} \sqrt{p^2_F + \frac{{\bf k}^2}{4} \sin^2{\phi}}-\frac{{\bf k}^2}{2} \cos^2{\phi}} \nonumber \\
&& \times \frac{p_F^2}{m\omega_k  + |{\bf k}| \cos{\phi} \sqrt{p^2_F - \frac{{\bf k}^2}{4} \sin^2{\phi}} + \frac{{\bf k}^2}{2} \cos^2{\phi}} . 
\label{l_2_1}
\end{eqnarray}
For $|{\bf k}| > 2p_F$ the expression for $\Pi_{pp}$ is more complex, but we do not need it as in this region the pole and branch cut are in the same half-plane of $\omega_k$ and the frequency integral vanishes.
  
Substituting (\ref{l_2}) into (\ref{l_1}), evaluating the frequency 
integral over a half plane where there is no double pole 
(this requires separate considerations for $|{\bf k}| > 2p_F \cos \theta_k$ and $|{\bf k}|< 2p_F \cos \theta_k$), and evaluating the remaining integrals over $\phi$,
 $\theta_k$ and $|{\bf k}|$ numerically, we obtain
 \begin{equation}
\delta{\tilde\Sigma}_1(\omega,\epsilon_{\mathbf{p}})= C \left(\frac{mU}{2\pi}%
\right)^2\left(\omega - \epsilon_{\mathbf{p}}\right)+\epsilon_{\mathbf{p}%
}\left(\frac{mU}{2\pi}\right)^2.  \label{new_n3}
\end{equation}
where, as before, $C = 0.6931\dots =\ln 2$.

 The total particle-particle self-energy, given by the sum of 
Eqs.~(\ref{new_n2}) and (\ref{new_n3}), is indeed the same as the total
particle-hole self-energy, given by the sum of Eqs.~(\ref{8-1}) and (\ref{8}%
), and total mass renormalization is the same as in Eq.~(\ref{4-1}). 
 However,  we see that mass renormalization now comes from both high and low energies.

\subsection{Momentum-dependent interaction}

For a momentum-dependent interaction $U(|\mathbf{k}|)$, the difference
between the diagrammatic and FL  formulas for the self-energy  is less
drastic. In particular, for a weak but momentum-dependent interaction, mass
renormalization to first order in $U(|\mathbf{k}|)$, as defined by Eq. (\ref
{n_4}), comes only from fermions on the FS.  

 For completeness, we also present perturbative
 results for $m^*/m$ and $Z$ for the 
 Coulomb interaction. At small $r_s$, both $m^*/m$ and $Z$ are quasi-linear in $r_s$ for small $r_s$.
In 3D,  $m^*/m = 1 -
(r_s/2 \pi)(4/9\pi)^{1/3} \ln {r^{-1}_s}$ and $Z = 1 -0.17696 r_s$ 
(Ref.~\onlinecite{gal}).  In 2D, $m^*/m =1 - (r_s/\pi\sqrt{2}) \ln {r^{-1}_s}$
 (Ref.~ \onlinecite{gal})
 and   $Z = 1 -  (r_s/\sqrt{2}) (1/2 + 1/\pi)$ (Ref.~\onlinecite{loss}).
The case of Coulomb interaction is special in that $Z$ comes from fermions in the  vicinity of the FS.

\section{Reconciliation of the Fermi-liquid and perturbative approaches}
\label{recons}

We have shown in the previous sections that  while mass renormalization comes from low energies in the FL theory, it 
generally contains both high- and low-energy contributions 
in a diagrammatic perturbation theory.
In Secs. ~\ref{sec:equiv}-\ref{sec:high}, we show how to reconcile the two approaches.

\subsection{Equivalence of the Fermi-liquid and perturbative aprpoaches for
a momentum-independent interaction}
\label{sec:equiv}

 To begin, we emphasize that the results for the
self-energy 
in  the FL  and perturbative approaches 
 need not to coincide identically, because $\Sigma_{%
\mathrm{FL}}$ given by Eq.~(\ref{15})  is only an expansion of the full
self-energy to first order in $\omega$ and $\epsilon_{{\bf p}}$, while $%
\Sigma_{\mathrm{pert}}$ given by Eqs.~(\ref{5},\ref{6_a}) and (\ref{6_b})
  contains all orders in $\omega$ and $\epsilon_{{\bf p}}$.
However, $\Sigma_{\mathrm{pert}}$ to first order in $\omega$ and $\epsilon_{{\bf p}}$ must coincide with $\Sigma_{\mathrm{FL}}$.

Comparing the two self-energies, 
 we see the difference:  while $\Sigma_{\mathrm{FL}}$, expressed via $%
\Gamma^\omega$,  contains both particle-hole and particles-particle bubbles, 
$\Sigma_{\mathrm{pert}}$  contains only a particle-hole bubble. We now show 
 that there exists a particular relation between the combinations of the
Green's functions which involve particle-hole and particle-particle
bubbles, namely 
\begin{equation}
\int \left(G_l G_{k-p_F+l} + G_l G_{k+p_F-l}\right)
\left(G_{k-\epsilon}-G_k\right) d_{kl} =0,  \label{10}
\end{equation}
where $\epsilon_{\mathbf{p}}$ is given by Eq.~(\ref{eps}). This relation is
valid to first order in $\omega$ and $\epsilon_{\mathbf{p}}$ and follows
from the identity,
 which we have already used implicitly when diagram {\it c}
in Fig.~\ref{fig:fig2}  was replaced by diagram {\it e}:
\begin{eqnarray}
&& \int d_{kl} G_l G_{k-p_F+l} \left(G_{k+\epsilon}-G_k\right) \notag  \\
&&=\int d_{kl} G_l G_{k+p_F-l} \left(G_{k-\epsilon}-G_k\right).  \label{9}
\end{eqnarray}
 The identity in Eq.~(\ref
{9})  is proven by relabeling the four-momenta, e.g., 
by relabeling the momenta in both terms 
in the second line as 
$k \to k-p_F + l + \epsilon$, and then relabeling $k+ \epsilon \to k$
 in the last term.
Equation (\ref{10}) follows from (\ref{9}) 
once $G_{k+
\epsilon}-G_k$ in the first line 
of Eq.~(\ref{9}) 
 is replaced by $G_{k}-G_{k-\epsilon}+ {\mathcal O}(\epsilon^2)$. 
 Adding Eq. (\ref{10}) to $\Sigma_{\mathrm{pert}} (\omega, \epsilon_{\mathbf{p%
}})$, we find, after simple algebra, that it becomes equal to $\Sigma_{%
\mathrm{FL}}$, i.e., the expressions for $m^*/m$ and $Z$ become exactly the
same as in Fermi liquid theory. Analyzing further the left-hand side of 
 Eq.~(\ref{10}), we find that both the particle-particle and particle-hole terms contain
 regular (high-energy) and  anomalous (low-energy) contributions.
Anomalous contributions contain only the single-particle dispersion $%
\epsilon_{\mathbf{p}}$, while regular contributions contain both $\epsilon_{%
\mathbf{p}}$ and $\omega$ terms. Expanding the regular contributions to
first order in $\omega$ and $\epsilon_{\mathbf{p}}$ and equating the
prefactors of $\epsilon_{\mathbf{p}}$ and $\omega$ terms, we obtain 
\begin{widetext}
\beq
 \omega \int d_{kl} \left(G_l G_{k-p_F+l} +  G_l G_{k+p_F-l}\right) G^2_k 
   =0  \label{11_a} 
\eeq
\beq \epsilon_{\mathbf{p}} \int d_{kl} \left(G_l G_{k-p_F+l} +  G_l G_{k+p_F-l}\right) G^2_k  \frac{ {\bf p}_F \cdot{\bf k}}{p^2_F}  
 =- \epsilon_{\mathbf{p}} \int  d_{kl}\left(G_l G_{k-p_F+l} +  G_l G_{k+p_F-l}\right) \delta G^2_k
 \frac{{\bf p}_F\cdot {\bf k} }{p^2_F}
\label{11_b}
\eeq   
\end{widetext}
where $\delta G^2_k$, defined by Eq.~(\ref{n_2}), projects the integral over 
$k$ onto the FS. 
 The left- and right-hand sides of Eq.~(%
\ref{11_b}) are anomalous (low-energy) and regular (high-energy)
contributions, respectively.

We see from Eq.~(\ref{11_a}) that  the addition of  Eq.~(\ref{10}) to $%
\Sigma_{\mathrm{pert}}$ does not change the result for the $\omega$ term in
a sense that there is no interplay between high-and low-energy
contributions, and Eq.~(\ref{11_a})  simply adds zero to the high-energy
contribution.  This explains why the $\omega$ terms in $\Sigma_{\mathrm{FL}}$
and $\Sigma_{\mathrm{pert}}$ are identical.  On the other hand, by adding
Eq. (\ref{10}) to $\Sigma_{\mathrm{pert}}$ we are changing the interplay
between the  high- and low-energy contributions to the $\epsilon_{\mathbf{p}}
$ terms. The regular contribution from Eq.~(\ref{11_b}) cancels the $%
\epsilon_{\mathbf{p}}$ term in $\delta \Sigma_1$, while  the anomalous
contribution renders $\delta \Sigma_2$ equal to the FL self-energy, Eq.~(\ref
{15}).

\subsection{Momentum-dependent interaction}

The results of the previous section can be readily extended to the case of a
momentum-dependent interaction.  In this situation,  we obtain, instead of
Eqs.~(\ref{3},\ref{6_a}) and (\ref{6_b}) 
\begin{widetext}
\bea
&&\Gamma^\omega_{\alpha\beta,\alpha\beta}  (p_F,q) = 
 U(0) +i \int    \frac{d^{D+1} l}{(2\pi)^{D+1}}
U^2 (
|\mathbf{p}_F-\mathbf{l}|)\left(G_l G_{q-p_F+l} + G_l G_{q+p_F-l}\right)
   \nonumber \\
&&-
\delta
_{\alpha\beta} \left[U(|\mathbf{q}-\mathbf{p}_F|) -i\int  \frac{d^{D+1} l}{(2\pi)^{D+1}}\left[ \left\{2 U(|\mathbf{q}-\mathbf{p}_F|) -  2 U(|\mathbf{q}-\mathbf{p}_F|) U(|\mathbf{p}_F-\mathbf{l}|)\right\} G_l G_{l+q-p_F} -
  U(|\mathbf{p}_F-\mathbf{l}|) U(|\mathbf{l}-\mathbf{q}|) G_l G_{q+p_F-l} \right]\right] \nonumber
\label{12}
\eea
and
$\delta\Sigma_{\mathrm{pert}}(\omega,\epsilon_{\mathbf{p}})=\Sigma_{\mathrm{pert}}(\omega,\epsilon_{\mathbf{p}})-\Sigma_{\mathrm{pert}}(0,0)=\delta\Sigma_1+\delta\Sigma_2$ with 
\bea
\delta\Sigma_1 (\omega,\epsilon_{\mathbf{p}}) &=&  
\int d_{lq}\left[2 G_l G_{k-p_F+l} \left\{U^2 (|\mathbf{q}-\mathbf{p}_F|) - 2 U(|\mathbf{p}_F-\mathbf{l}|) U(|\mathbf{q}-\mathbf{p}_F|)\right\}- 
 G_l G_{q+p_F -l} U(|\mathbf{p}_F-\mathbf{l}|) U(|\mathbf{l}-\mathbf{q}|)\right]
 \left(G_{k-\epsilon}-G_k\right)  \nonumber \\
\delta \Sigma_2 (\omega,\epsilon_{\mathbf{p}})& =& - \int d_{lq}
 \left[2 G_l G_{k-p_F+l} \left\{U^2 (|\mathbf{q}-\mathbf{p}_F|) - 2 U(|\mathbf{p}_F-\mathbf{l}|) U((|\mathbf{q}-\mathbf{p}_F|)\right\}- 
 G_l G_{q+p_F -l} U(|\mathbf{p}_F-\mathbf{l}|) U(|\mathbf{l}-\mathbf{q}|)\right] G^2_q\notag\\
&&\times \left(\omega - \epsilon_{\mathbf{p}} \frac{{\bf p}_F\cdot{\bf q}}{p^2_F}\right),
\label{14}
\eea
\end{widetext}
where we neglected first-order terms.

Comparing the  expressions for $\Gamma^{\omega}$ and $\delta\Sigma_{\mathrm{%
pert}}$, we see that $\Gamma^\omega$ [and, hence, $\Sigma_{\mathrm{FL}}$
given by Eq.~(\ref{15})]  again contains two extra terms not present in the
diagrammatic self-energy.  These two terms have the same overall factor of $%
U^2 (|\mathbf{p}_F-\mathbf{l}|)$.  After some re-arranging of the momenta 
in the products of three fermionic propagators, we  obtain  an analog of Eq.~(%
\ref{10}) for a momentum-dependent interaction as 
\begin{widetext}
\beq
\int  d_{ql}U^2 (|\mathbf{p}_F -\mathbf{l}|) \left(G_l G_{k-p_F+l} +  G_l G_{q+p_F-l}\right) 
\left(G_{k-\epsilon}-G_k\right)
\label{10_a}
\eeq
\end{widetext}
Adding this expression to $\delta\Sigma_{\mathrm{pert}}$,  we find after 
some algebra that high-energy contributions to $m^*/m$  cancel  and, to
first order in $\omega$ and $\epsilon_{\mathbf{p}}$, $\Sigma_{\mathrm{pert}}$
becomes equal to $\Sigma_{\mathrm{FL}}$.

\subsection{Higher orders of the perturbation theory}
\label{sec:high}

So far, we have focused only on the lowest-order perturbation theory. One
can show, however, that Eq.~(\ref{10_a}) remains valid if the bare fermionic
propagators are replaced by the 
 full ones and $U(|\mathbf{q}|)$ is replaced by a fully 
 renormalized interaction
which depends not only on the momentum but also on frequency. This is so
because  Eq.~(\ref{10_a}) is proven simply by re-arranging internal
four-momenta. Next, the full perturbative  self-energy 
 is also obtained from the second-order result, Eq. (\ref{14}), 
by dressing the interactions and propagators. Adding 
Eq. (\ref{10_a})  to the full perturbative self-energy, expressed via the
full
 propagators and full interactions, we immediately recover $\Sigma_{\mathrm{FL%
}}$ simply because the previous proof of this statement 
did not 
rely on the specific 
forms of $G$ and $U$.

\section{$SU(N)$-invariant Fermi liquid}

\label{largeN}

\subsection{$SU(N)$ vs $SU(2)$}

\label{sunsu2} In this section, we discuss the interplay between the FL- and
perturbation theories for a system of interacting fermions with a large
number of flavors $N$. Fermi-liquid properties of such a system were
discussed both in terms of RG~\cite{chitov} and pertubation theory for the
case of a Coloumb interaction.~\cite{iordan,suhas} In the limit when the
 $N$ times a (dimensionless) coupling constant  is larger than one, renormalization of not only the $Z$ factor
but also of the effective mass comes from energies much higher than the
Fermi energy. Based on the observation, the authors of Ref.~%
\onlinecite{iordan} argued that an $SU(N)$-invariant FL is not of the same
type as discussed in the framework of the Landau theory. We show here that
this is not the case: the FL- and perturbation theories give the same
results for the $SU(N)$-invariant case as well. To see this, however, one
needs to collect next-to-leading terms in the large-$N$ expansion of the FL
theory, whereas the perturbation theory can be evaluated only to the leading
order in $1/N$.

The difference between the large-$N$ expansions for the FL- and perturbation
theories is most dramatic for the case of 
 a momentum-independent
interaction in 2D, 
and for brevity we consider only this case
here.
 In the $SU(N)$ case, each particle-hole bubble is multiplied by a factor of $N$. The second-order
self-energy contains two diagrams --\textit{c} and \textit{d} in Fig.~\ref
{fig:fig2}--the first of which acquires a factor of $N$ while the second does
not. Consequently, the perturbative effective mass acquires a factor of $N-1$
compared to the result in Eq.~(\ref{4-1}) 
\begin{equation}
\frac{m^*}{m}=1+\frac{N-1}{2}\left(\frac{mU}{2\pi}\right)^2\approx 1+\frac{N%
}{2}\left(\frac{mU}{2\pi}\right)^2,  \label{4-1N}
\end{equation}
where the last result applies to the large-$N$ limit. On the other hand, the
only diagram for $\Gamma^{\omega}$ which acquires a factor of $N$ --diagram 
\textit{f} in Fig.~\ref{fig:fig1}--does not depend on the angle between the
initial fermionic momenta $\mathbf{p}_F$ and $\mathbf{q}_F$, because the static
particle-hole bubble is independent of the momentum in 2D for $|\mathbf{p_F}-%
\mathbf{q_F}|\leq 2p_F$. Therefore, the leading term in the $1/N$ expansion
for $\Gamma^{\omega}$ does not contribute to mass renormalization. To
resolve this contradiction, one needs to recall that, when deriving the FL
result for $m^*$ (\ref{1_b}) we divided the trace of $%
\Gamma^{\omega}$ by the number of spin components. This is the origin of the
factor of two in the prefactor $A_D$.  
 The formula for the $SU(N)$
case is obtained from Eq.~(\ref{1_b}) simply by replacing $2$ in $A_D$ by $N$; in 2D,
we have 
\begin{equation}
\frac{1}{m^*} = \frac{1}{m} - \frac{Z^2}{N(2\pi)^2} \sum_{\alpha\beta} \int
\Gamma^\omega_{\alpha\beta,\alpha\beta} (p_F,q_F) \frac{\mathbf{p}_F\cdot 
\mathbf{q}_F}{p^2_F}~ d \Omega_q.  \label{1_bN}
\end{equation}
Diagrams for \lq\lq direct\rq\rq\/ processes ($\mathbf{p}\to \mathbf{p},\mathbf{q}\to 
\mathbf{q}$) enter $\Gamma^{\omega}$ with a factor of $\delta_{\alpha\gamma}\delta_{\beta,%
\delta}$, which becomes equal to one for $\gamma=\alpha$ and $\delta=\beta$.
Therefore, the trace of the direct contribution to $\Gamma^{\omega}$ gives a
factor of $N^2$ which, upon dividing by an overall factor of $N$ in Eq.~(\ref
{1_bN}), gives an ${\mathcal O}(N)$ contribution to $m^*$. To obtain an
${\mathcal O}(N)$ term in $m^*$ from the FL theory, one thus needs to collect 
\textit{all direct} ${\mathcal O}(1)$ diagrams for $\Gamma^{\omega}$. On the
other hand, diagrams for \lq\lq exchange\rq\rq\/ processes ($\mathbf{p}\to \mathbf{q},%
\mathbf{q}\to \mathbf{p}$) enter $\Gamma^{\omega}$ with a factor of  $\delta_{\alpha\delta}%
\delta_{\beta\gamma}$, which becomes equal to $\delta_{\alpha\beta}$ for $%
\gamma=\alpha$ and $\delta=\beta$. The trace of the exchange contribution is
of order $N$, which translates into a subleading, ${\mathcal O}(1)$ term in $m^*$.
Therefore, one can neglect exchange processes in $\Gamma^{\omega}$ in the
large-$N$ limit. A physical reason for this simplification is obvious: since
the large-$N$ limit is inherently semiclassical, the Pauli principle becomes
irrelevant.

The same conclusion also follows from an identity which involves the
generators ${\hat T}$ of the $SU(N)$ group~\cite{chitov} 
\begin{equation}
\delta_{\alpha\delta}\delta_{\beta\gamma}=\frac{1}{2}\sum^{N^2-1}_{a=1}{\hat
T}^a_{\alpha\gamma}{\hat T}^a_{\beta\delta}+\frac{1}{N}\delta_{\alpha\gamma}%
\delta_{\beta\delta}.
\end{equation}
The delta-symbols on the left (right) occur in exchange (direct)
contributions to $\Gamma^{\omega}$, correspondingly. Since ${\hat T}^a$ are
traceless, it follows immediately that the trace of the exchange contribution to $\Gamma^{\omega}$ is by a
factor of $N$ smaller than the trace of the direct contribution.

Coming back to the second-order $\Gamma^{\omega}$ for arbitrary $N$, we need
to consider all subleading, ${\mathcal O}(1)$ diagrams in Fig.~(\ref{fig:fig1}). Diagrams \textit{e}
and \textit{g} contain particle-hole bubbles evaluated at $|\mathbf{p}_F-%
\mathbf{q}_F|$ and, therefore, do not contribute to mass renormalization.
Direct particle-particle diagram \textit{c} contributes an 
$
{\mathcal O}(N)$ term 
to $m^*$, while its exchange counterpart \textit{d} contributes an ${\mathcal O}(1)$. 
A combined contribution of diagrams \textit{c} and \textit{d}
is $(N-1)$ times the $SU(2)$ result, which is the same as in Eq.~(\ref{4-1N}%
).

In the $SU(2)$ case, the perturbative regime implies that the interaction is
weak in a sense that $mU\ll 1$. In the $SU(N)$ case with a large number of
flavors, there is an intermediate range of interactions, defined by the
condition $1/N\ll Um\ll 1$, where the perturbation 
 theory can be resummed to infinite order. It is instructive to compare the perturbation theory with
the FL formalism in this case. 
 The subsequent analysis will be performed in the Matsubara technique at $T=0$.

\subsection{Pertubation theory for the self-energy  in the regime
 $1/N\ll  Um
 \ll
  1$}

\begin{figure}[tbp]
\label{fig:fig3}\centering
\includegraphics[width=1.0\columnwidth]{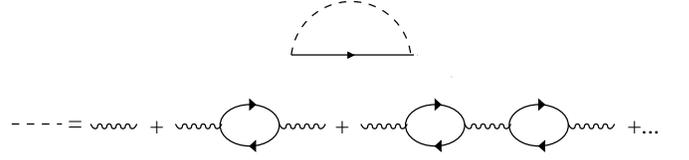}
\caption{ Self-energy in the large-$N$ limit. The dashed line is an effective interaction given by Eq.~(\ref{rpa}).}
\end{figure}

In the large-$N$ limit, the perturbative self-energy is given diagrams with a
maximal number of the particle-hole bubbles at each order in $U$. The sum of
such diagrams is equivalent to a first-order diagram, shown in Fig.~\ref{fig:fig3}, 
\begin{equation}
\Sigma_{\mathrm{pert}}(\omega,\epsilon_{\mathbf{p}})=\int \frac{d^2 k}{%
(2\pi)^2}\int\frac{d\omega_{k}}{2\pi} G\left(\omega+\omega_{k},\mathbf{p}+%
\mathbf{ k}\right) {\tilde U}\left(\omega_{k},\mathbf{ k}\right),  \label{sigmaN}
\end{equation}
where the effective interaction is of an RPA form
\begin{equation}
{\tilde U}\left(\omega_{k},\mathbf{ k}\right)=\frac{U}{1-NU\Pi_{\mathrm{ph}%
}\left(\omega_{k},\mathbf{ k}\right)}.  \label{rpa}
\end{equation}
As in the $SU(2)$ case, the self-energy can be separated into anomalous and
regular part. The anomalous part is similar to Eq.~(\ref{s2}), except for
the second-order effective interaction $U^2\Pi_{\mathrm{ph}}\left(\omega_{k}=0,%
\mathbf{ k}\right)$ is replaced by ${\tilde U}\left(\omega_{k}=0,\mathbf{ k}%
\right)$. Still, as $\Pi_{\mathrm{ph}}\left(\omega_{k}=0,\mathbf{ k})\right)$
is independent of $|\mathbf{ k}|$ for $|\mathbf{ k}|\leq 2p_F$, the integral
over $|\mathbf{ k}|$ vanishes, and the anomalous part of the self-energy does
not renormalize the effective mass.

In the regular part of the self-energy, we assume--and then justify--that
the limit of $NUm\gg 1$ corresponds to large momentum transfers: $|\mathbf{ k}%
|\gg p_F$. In this limit, the particle-hole bubble becomes
\cite{iordan} 
\begin{equation}
\Pi_{\mathrm{ph}}\left(\omega_{k},\mathbf{ k}\right)=-\frac{p_F^2}{2\pi}\frac{%
E_{\mathbf{ k}}}{\omega^2_q+E_{\mathbf{ k}}^2},  \label{bubble_large_q}
\end{equation}
where $E_{\mathbf{ k}}=\mathbf{ k}^2/2m$. At fixed number density of particles 
$n$, the area of the Fermi surface is inversely proportional to the
number of flavors 
\begin{equation}
p_F^2/4\pi=n/N.
\label{pauli}\end{equation}
With this normalization, the product $N
%
\Pi_{\mathrm{ph}}\left(\omega_{k},%
\mathbf{ k}\right)=-2nE_{\mathbf{ k}}/(\omega^2_{k}
+E_{\mathbf{ k}}^2)$ is
independent of $N$. A pole of ${\tilde U}$ in real frequencies corresponds
to the collective (zero-sound) mode with dispersion $\omega_{k}=\sqrt{E_{%
\mathbf{ k}}^2+E_0E_{\mathbf{ k}}}$, which interpolates between sound-like
excitations for $E_{\mathbf{ k}}\ll E_0\equiv 2nU$ and particle-like
excitations for $E_{\mathbf{ k}}\gg E_0$. Note that $E_0/E_F\sim NUm\gg 1$.

According to Eq.~(\ref{a}), mass renormalization is determined by a
stand-alone $\epsilon_{\mathbf{p}}$ term in the self-energy. This term is
obtained by expanding the single-particle dispersion in the argument of the
Green's function in Eq.~(\ref{sigmaN}) as $\epsilon_{\mathbf{p}+\mathbf{ k}%
}=\epsilon_{\mathbf{p}}+\epsilon_{\mathbf{p}}\mathbf{p}_F\cdot \mathbf{ k}%
/p_F^2 +\mathbf{p}_F\cdot \mathbf{ k}/m+E_{\mathbf{ k}}$ and differentiating
the regular part of the self-energy with respect to the second term in $%
\epsilon_{\mathbf{p}+\mathbf{ k}}$ at $\epsilon_{\mathbf{p}}=\omega=0$. This
yields 
\begin{equation}
\frac{m}{m^*}=1-\int \frac{d^2 k}{(2\pi)^2}\int\frac{d\omega_{k}}{2\pi} G\left(%
\mathbf{p}_F+\mathbf{ k},\omega_{k}\right) {\tilde U}\left(\omega_{k},\mathbf{ k}%
\right)\frac{\mathbf{p}_F\cdot\mathbf{ k}}{p_F^2}.  \label{mN1}
\end{equation}
Since, according to our 
 assumption, 
large $|\mathbf{ k}|$ control the
integral in Eq.~(\ref{mN1}), the $E_{\bf k}$ term in the denominator of Green's function is larger than the ${\bf p}_F\cdot {\bf k}$ term.
Expanding the Green's function to first order in $\mathbf{p}_F\cdot \mathbf{ k}$, integrating over the
angle,
  and switching from integration over $|\mathbf{ k}|$ to integration over 
$E_{\mathbf{ k}}$, we obtain 
\bea
\frac{m}{m^*}&=&1-\frac{Um}{2\pi^2} \int^{\infty}_0 dE_{\mathbf{ k}%
}\int^{\infty}_{-\infty} d\omega_{k} \frac{\omega_{k}^2+E_{\mathbf{ k}}^2}{%
\omega_{k}^2+E_{\mathbf{ k}}^2+E_0E_{\mathbf{ k}}}\notag\\
&&\times\frac{E_{\mathbf{ k}}}{%
(i\omega_{k}-E_{\mathbf{ k}})^3}.
\label{mN3}
\eea
Now it is obvious that typical $|\omega_{k}|\sim E_{\mathbf{ k}}\sim E_0\gg E_F$%
, which justifies our original assumption. Performing remaining integrations, we finally obtain 
\begin{equation}
\frac{m^*}{m}=1+\frac{mU}{8\pi}.  \label{mN2}
\end{equation}
As it is also the case for the Coulomb interaction,~\cite{iordan,suhas} the effective mass in the $NUm\gg 1$ limit does not
depend on $N$. Note that Eq.~(\ref{mN2}) is valid only for a repulsive and
sufficiently weak interaction ($Um\ll 1$), so that mass renormalization is
still a small albeit non-perturbative effect. Notice also that the normalization condition
(\ref{pauli}) was not essential: if it is not imposed, the high-energy scale $E_0$ is replaced by
${\tilde E}_0=(NUm/\pi)E_F\gg E_F$. However, since the high-energy scale drops out from the formula (\ref{mN3})
for the effective mass, changing $E_0$ by ${\tilde E}_0$ does not affect the result for $m^*$.

Differentiating the regular part of the self-energy with respect to $i\omega$
and performing integrations in a way similar to the effective mass case, we
obtain the $Z$ factor in the $NUm\gg 1$ limit 
\begin{equation}
Z=1-\frac{mU}{4\pi}.
\end{equation}

\subsection{Fermi-liquid formalism in the large-$N$ limit}
\begin{figure}[tbp]
\centering
\includegraphics[width=1.0\columnwidth]{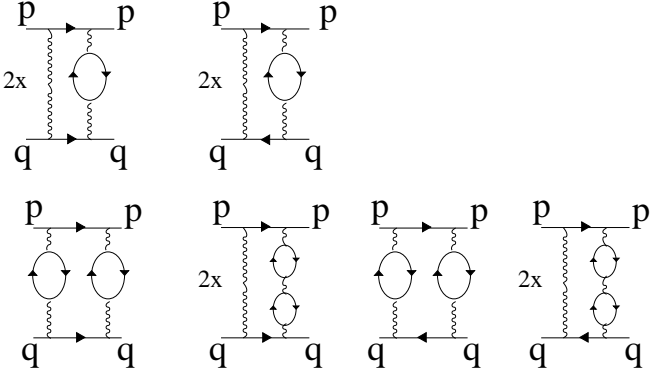}
\caption{ Diagrams for the Fermi-liquid vertex $%
\Gamma^{\protect\omega}_{\protect\alpha\protect\beta,\protect\gamma\protect%
\delta}(p,q)$ in the large-$N$ limit and to fourth order in the momentum-independent interaction $U=\mathrm{const}$. We select only those diagrams which contribute
to mass renormalization in 2D. Note that particle-particle and particle-hole diagrams differ by the directions of arrows on bottom fermionic lines.}
\label{fig:fig4}
\end{figure}

As we explained in Sec.~\ref{sunsu2}, to obtain the effective mass in the FL
formalism, one needs to collect all \textit{direct} diagrams for $\Gamma
^{\omega }$ to next-to-leading order in $1/N$. This arduous task is
simplified dramatically in the case of a momentum-independent interaction in
2D, where diagrams with particle-hole bubbles at $\mathbf{p}-\mathbf{q}$ do
not contribute to mass renormalization. To second order in $U$, the leading
order in $N$ is $N^{1}=N$ and next-to-leading order is $N^{0}=1$. There is
only one direct, $\mathcal{O}\left( N^{0}\right) $ diagram that does not
contain $\Pi _{\mathrm{ph}}\left( \omega =0,\mathbf{p}-\mathbf{q}\right) $--diagram \textit{c} in Fig.~\ref{fig:fig1}. To third order in $U$, there
are only two inequivalent diagrams of order $\mathcal{O}\left( N^{1}\right)
$, shown in Fig. \ref{fig:fig4}. One of them renormalizes the second-order
particle-particle diagram \emph{c}, while the other renormalizes the
particle-hole diagram \emph{e } in Fig. \ref{fig:fig1}. Note that the
particle-hole bubbles in both diagrams are integrated over internal
four-momenta and, hence, do contribute the angular dependence of $\Gamma^{\omega}$. To fourth order in $U$, there are
four inequivalent diagrams, also shown in Fig. \ref{fig:fig4}, etc. It is easy to see
that the overall combinatorial coefficients  for particle-particle and
particle-hole diagrams 
of order $U^{n}$ 
are both equal to $n-1$. 
 Collecting all orders, we
obtain for the angular-dependent part of $\Gamma ^{\omega }$%
\begin{widetext}
\begin{eqnarray}
\Gamma _{\alpha \beta ,\gamma \delta }^{\omega }\left( p_{F},q_{F}\right) 
&=&-U^{2}\delta _{\alpha \beta}\delta_{\gamma \delta }\int \frac{d^{3}l}{\left(
2\pi \right) ^{3}}\left[ G_{p+l}G_{q+l}+G_{p+l}G_{q-l}\right]
\sum_{n=0}^{\infty }\left( n+1\right) \left[ NU\Pi _{\mathrm{ph}}\left(
l\right) \right] ^{n}  \notag \\
&=&-U^{2}\delta _{\alpha \beta} \delta_{\gamma \delta }\int \frac{d^{3}l}{\left(
2\pi \right) ^{3}}\left[ G_{k+l}G_{q+l}+G_{k+l}G_{q-l}\right] \frac{1}{\left[
1-NU\Pi _{\mathrm{ph}}\left( l\right) \right] ^{2}},  \label{gN1}
\end{eqnarray}
\end{widetext}
where we added the ${\mathcal O}(U^{2})$ particle-hole diagram, which does not contribute to
mass renormalization, to the right-hand side of Eq. (\ref{gN1}). 
 The result
for  $\Gamma _{\alpha \beta ,\gamma \delta }^{\omega }$ is greatly
simplified in the limit of $NUm\gg 1.$ As in the previous section, we
replace $\Pi _{\mathrm{ph}}\left( l\right) $ by its large-momentum
asymptotic form (\ref{bubble_large_q}) and expand the products of Green's functions as 
\begin{eqnarray}
G_{p+l}G_{q+l} &=&\frac{\left( v_{F}l\right) ^{2}}{\left( i\omega
_{l}-E_{\mathbf{l}}\right) ^{4}}\cos \theta _{pl}\cos \theta _{ql}+\dots\notag  \\
G_{p+l}G_{q-l} &=&-\frac{\left( v_{F}l\right) ^{2}}{\left( \omega
_{l}^{2}+E_{\mathbf{l}}^{2}\right) ^{2}}\cos \theta _{pl}\cos \theta
_{ql}+\dots ,
\end{eqnarray}
where $ \dots $ stand for angular-independent and higher order
terms, $\theta _{k_{1}k_{2}}\equiv \angle \left( \mathbf{k}_{1}\mathbf{,k}%
_{2}\right)$, and, as before,  $E_{\mathbf{l}}=l^{2}/2m$. After trivial angular
integration and some simplifications, we obtain
\begin{widetext}
\bea
\Gamma _{\alpha \beta ,\gamma \delta }^{\omega }\left( p_{F},q_{F}\right)
&=&-\delta _{\alpha \beta} \delta_{\gamma \delta }\frac{U^{2}k_{F}^{2}}{\pi ^{2}}\cos
\theta _{pq}\int^{\infty}_0 dE_{\mathbf{l}}\int^{\infty}_{-\infty} d\omega _{l}\frac{i\omega _{l}E_{%
\mathbf{l}}^{2}}{\left( i\omega _{l}-E_{\mathbf{l}}\right) ^{2}}\frac{1}{%
\left( \omega _{l}^{2}+E_{\mathbf{l}}^{2}+E_{0}E_{\mathbf{l}}\right) ^{2}}\notag\\
&&=\delta _{\alpha \beta} \delta_{\gamma \delta }\cos\theta_{pq}\frac{U}{2N}.\label{gN2}
\eea
\end{widetext}
Substituting Eq.~(\ref{gN2}) into the formula for the effective mass (\ref{4-1N}), we reproduce the result of the perturbation theory, Eq.~(\ref{mN2}).

To reproduce the perturbative result for $Z$, one needs to evaluate 
$\Gamma _{\alpha \beta ,\gamma \delta }^{\omega }\left( p_{F},q\right)$ for $q$ away from the FS. We did not attempt to do this.   

\section{Conclusions}

\label{conc}
In this paper, we analyzed an interplay between high- and low-energy contributions to two fundamental Fermi-liquid parameters--
the quasiparticle $Z$-factor and effective mass $m^*$--obtained in two different ways: via a general FL formalism and via a diagrammatic perturbation theory. 
In both cases, $Z$ and $m^*$ are extracted from the fermionic self-energy. 
 In the FL formalism, the self-energy $\Sigma_{\mathrm{FL}}$ is obtained 
from the Pitaevskii identities for the derivatives of the Green's functions
(following from the partcile number conservation and Galilean invariance)
and expressed via  an antisymmetrized FL vertex $\Gamma^\omega$.  In the perturbation theory, 
 the self-energy $\Sigma_{\mathrm{pert}}$  is obtained 
 in series of non-antisymmerized interaction $U(|{\bf k})|$.
 To any order in $U(|{\bf k}|)$,  the two self-energies are not
identical when expressed in terms of fermionic Green's functions, but 
 certainly yield
the same expressions for 
 $m^*/m$ and $Z$. 
We found, however, that identical results for $m^*$ in the two approaches are determined by 
 different regions of energies.
Whereas the FL-theory $m^*$ comes from low energies, i.e., from the vicinity of the Fermi surface, the perturbative $m^*$ includes, in general, contributions from both low- and high energies.
 Only the sum of the two contributions coincides with the FL result for $m^*$.
 We found that the equivalence of 
$m^*/m$ in the
 two approaches is based 
on a particular identity 
 for the products of
fermionic Green's functions, Eq.~(\ref{10_a}) which relates 
 the low- and high-energy contributions to the effective mass. 
On the other hand, renormalization of $Z$
comes only from high-energy fermions in both approaches.
 We obtained  the expression for $Z$ in a 2D Fermi liquid with short-range interaction.  

We also analyzed the difference between the FL- and perturbative approaches for a system of interacting fermions with $SU(N)$ symmetry in the limit
of $N\gg 1$. We showed that 
 mass renormalization  in the diagrammatic formalism 
  comes from high energies, and that  equivalent expressions for the effective mass are obtained  
 in the two formalisms 
 only if one collects next-to-leading terms in the $1/N$ expansion for $\Gamma^{\omega}$. We
 obtained a closed expression for $\Gamma^{\omega}$ for the case of momentum-independent interaction in 2D.

The lesson to be learned from this consideration is that one has to be careful 
 with
eliminating  high-energy fermions from the problem. 
 While it is tempting to
reduce the problem to 
 that of
 low-energy fermions with an effective interaction  and
consider only the low-energy contribution to $m^*/m$, 
 this would give an incorrect
result for $m^*$. The reason is that, 
 in the process of integrating out high-energy fermions,
the quasiparticle mass changes from its bare value, $m$, to a new one, $m_B$. The
difference $m_B/m -1$ comes from high energies.  Only the 
 the combined effect of 
 high-energy renormalization,
 which replaces $m$ 
 by $m_B$, and low-energy renormalization, which involves onlt the Fermi-surface states, yields the agreement with the FL theory.

\section{Acknowledgement}

We acknowledge helpful discussions with  E. Fradkin and R. Shankar. This
work was supported by  NSF-DMR--0906953 (A. V. Ch.) and NSF-DMR-0908029
(D. L. M.). The authors acknowledge the 
support from MPIPKS, Dresden, where
a part of this work 
was done. 
A.V. Ch acknowledges the
 hospitality of the
Aspen Center for Physics.


\begin{thebibliography}{99}
\bibitem{agd}  A.\ A.\ Abrikosov, L.\ P.\ Gorkov, and I.\ E.\ Dzyaloshinski, 
\emph{Methods of quantum field theory in statistical physics}, (Dover
Publications, New York, 1963); E. M. Lifshitz and L. P. Pitaevski, \emph{%
Statistical Physics}, (Pergamon Press, 1980).

\bibitem{pines}  D. Pines and P. Nozieres, {\it The theory of quantum liquids},
(Addison-Wesley, Menlo Park, 1966).

\bibitem{anderson}  Anderson P.W. {\it Basic notions of Condensed Matter Physics}%
, (Benjamin-Cummings, Melno Park, 1984).

\bibitem{baym}  G. Baym and C. Pethick, {\it Landau Fermi Liquid theory}, (Wiley,
New York, 1991).

\bibitem{pit}  L. P. Pitaevskii, Sov. Phys. JETP {\bf 10}, 1267 (1960).


\bibitem{abrikosov58}  A. A. Abrikosov and I. M. Khalatnikov, Sov. Phys.
JETP \textbf{10}, 132 (1958);

\bibitem{galitskii}  V.M. Galitskii, Sov. Phys. JETP \textbf{7}, 104 (1958).

\bibitem{shankar}  R. Shankar, Rev. Mod. Phys. \textbf{66}, 129 (1994).


\bibitem{comment_Z}  To the best of our knowledge, the second-order result
for $Z$ [Eq.~(\ref{4_aa})] was not derived in the literature within the FL
theory, i.e., using Eq.~(\ref{1_a}).

\bibitem{gal}  See, e.g., V. M. Galitski and S. Das Sarma, Phys. Rev. B 
\textbf{70}, 035111 (2004) and references therein.

\bibitem{loss}  G. Burkard, D. Loss, and E.V. Sukhorukov, Phys. Rev. B 
\textbf{61} R16303 (2000).

\bibitem{randeria}  J. R. Engelbrecht, M. Randeria, and L. Zhang, \prb {\bf
45}, R10135  (1992).

\bibitem{chitov} G. Y. Chitov and D. S{\'e}n{\'e}chal, \prb {\bf 52}, 13847 (1995).

\bibitem{iordan} S. V. Iordanskii and A. Kashuba, JETP Lett. {\bf 76}, 563 (2002).

\bibitem{suhas} S. Gangadharaiah and D. L. Maslov, Phys. Rev. Lett. {\bf 95}, 186801 (2005).  
\end{thebibliography}
\end{document}